\documentclass[prd,aps,twocolumn,a4paper,showkeys,nofootinbib]{revtex4-2}

\usepackage{graphicx,psfrag}
\usepackage{mathrsfs}
\usepackage{amsmath,amsfonts,amssymb}
\usepackage{multirow}
\usepackage{comment}
\usepackage{subfigure}
\usepackage{appendix}
\usepackage{hyperref}
\usepackage{ragged2e}
\usepackage{enumitem}
\usepackage[percent]{overpic}
\usepackage{float}

\newcommand{\be}{\begin{equation}}
\newcommand{\ee}{\end{equation}}
\newcommand{\bea}{\begin{eqnarray}}
\newcommand{\eea}{\end{eqnarray}}
\newcommand{\bel}{\begin{align}}
\newcommand{\eel}{\end{align}}

\def\GMc2{{\rm G M_{\odot} c^{-2}}}

\def\kt2{\kappa^\text{T}_2}

\usepackage{pifont} % \cmark \xmark

\usepackage{color}
\definecolor{cyan}{rgb}{0,0.9,0.9}
\definecolor{orange}{rgb}{0.9,0.5,0}
\definecolor{magenta}{rgb}{1,0,1}
\definecolor{purple}{rgb}{0.8,0.4,0.8}
\definecolor{gray}{rgb}{0.8242,0.8242,0.8242}

\begin{document}

\title{Controlling artificial surface heating in neutron star simulations: Application to hybrid equations of state}
%Controlling Artificial Stellar Surface Heating in Hybrid Neutron Star Equations of State

\author{Georgios \surname{Doulis}}

\affiliation{Institut f{\"u}r Theoretische Physik, Goethe-Universit{\"a}t Frankfurt, 60438 Frankfurt am Main}

\date{\today}

\begin{abstract}
The treatment of the stellar surface in binary neutron star simulations is crucial for the accuracy of the numerical evolution and the physical reliability of the predicted observables. Numerical artifacts associated with the treatment of steep gradients near the stellar surface can produce spurious heating during the inspiral, leading to an artificial increase of the internal energy and an unphysical expansion of the neutron star. In this work, we investigate the effectiveness of the entropy-based flux-limiting (EFL) scheme in mitigating these numerical effects within the finite-difference code BAM. We perform simulations of both isolated neutron stars and binary neutron star inspirals employing a representative set of hybrid equations of state. We show that the EFL scheme significantly reduces artificial surface heating. This reduction is observed consistently across all stellar models and binary configurations considered, demonstrating that the reduction of numerical heating is a robust feature of the EFL method. 
\end{abstract}


\maketitle

\section{Introduction}

Neutron stars constitute a unique astrophysical laboratory for exploring matter under extreme conditions. 
%(, combining supranuclear densities, strong-field gravity and intense magnetic fields)
Their macroscopic properties and long-term evolution encode valuable information about the equation of state (EoS) of dense matter and the underlying microphysical processes governing neutron star (NS) transport properties, composition and radiative behaviour \cite{Lattimer:2004pg,Haensel:2007yy}. As many of these regimes are inaccessible to analytic treatment, numerical simulations have become an essential tool for studying NS structure, dynamics and evolution \cite{Baiotti:2016qnr}.
%across a wide range of astrophysical scenarios, including isolated NSs, accreting systems and binary neutron star (BNS) mergers \cite{Baiotti:2016qnr}.

Despite significant advances in numerical relativity \cite{Read:2013zra,Radice:2013hxh,Haas:2016cop,Most:2019kfe}, accurate modelling of neutron stars remains challenging, particularly in the vicinity of the stellar surface. The surface region is characterized by sharp gradients in the hydrodynamical variables, marking the transition from dense stellar matter to a tenuous exterior. In binary systems, the stellar surface is advected across the grid and undergoes continuous tidal deformations, thereby exacerbating the challenges faced by the high-resolution shock-capturing (HRSC) schemes \cite{Toro:1999} used in the evolution of the fluid. These effects lead to an inaccurate representation of the stellar boundary, with consequences for the simulation as a whole \cite{Doulis:2022vkx,Gittins:2024jui}.

One direct consequence of the inaccurate numerical treatment of the NS surface is the appearance of artificial surface heating, i.e. a spurious increase of the thermal energy that is not associated with any physical heating mechanism. Although the heating originates at the surface, the temperature increase rapidly spreads into the stellar interior as the simulation progresses \cite{Gittins:2024jui}. Such artificially elevated temperatures can bias comparison between simulations and observations, impact inferred constraints on NS microphysics and contaminate parameter estimation. This calls for the development of numerical strategies that mitigate artificial surface heating and maintain physically consistent NS temperatures during the inspiral phase.

In \cite{Gittins:2024jui} it was convincingly demonstrated that artificial heating alters the structure and the tidal properties of NSs in BNS merger simulations. Although the analysis presented there focused on finite-temperature EoSs, similar effects (not to the extent presented in \cite{Gittins:2024jui}) should also be expected in simulations employing hybrid EoSs \cite{Shibata:2005ss} as the underlying numerical mechanisms responsible for artificial heating are not specific to a particular thermodynamic treatment. 

In the present work we investigate the performance of the fast-converging entropy-based flux-limiting scheme developed in \cite{Doulis:2022vkx} in controlling artificial heating in simulations employing hybrid EoSs. The EFL scheme builds upon the flux limiter proposed in \cite{Guercilena:2016fdl}, which pioneered the use of entropy in the construction of flux limiters. The use of hybrid EoSs in the present study is motivated by the fact that, to our knowledge, the EFL method has so far been implemented only in the finite-difference code BAM \cite{Bruegmann:2003aw,Bruegmann:2006ulg}. Since BAM does not evolve the temperature explicitly, and the finite-temperature extension of the code does not employ the EFL scheme \cite{Gieg:2022mut,Schianchi:2023uky}, hybrid EoSs are employed to model thermal effects in NS matter while using the EFL formulation. In this setup, thermal effects (and in particular artificial heating) are encoded in the evolution of the specific internal energy, which we adopt as the primary diagnostic tool to monitor artificial heating throughout this work.

Although artificial surface heating is expected to be less pronounced in simulations employing hybrid EoSs than in those using fully finite-temperature treatment, the resulting temperature increase should be detectable in both isolated and binary neutron star simulations. Identifying and quantifying this effect in the hybrid case should provide a controlled testing ground for assessing the ability of numerical strategies to mitigate artificial heating. In particular, evidence that the EFL scheme can effectively control surface heating in simulations with hybrid EoSs would strongly suggest that similar behaviour may be expected in finite-temperature simulations once the method is implemented in codes that evolve the temperature explicitly.

The article is organised as follows. Sec.~\ref{sec:num_setup} introduces the numerical methods and the EFL scheme, together with the hybrid EoSs and the setup used for the simulations. In Sec.~\ref{sec:sns}, we investigate artificial heating in isolated NS configurations and assess the performance of the EFL scheme using a set of TOV models. In Sec.~\ref{sec:bns}, we extend the analysis to BNS inspirals and examine the impact of artificial heating in fully dynamical simulations. Finally, in Sec.~\ref{sec:conclusions}, we summarize our main findings and discuss possible directions for future work.

Throughout this work we use geometric units. We set $c = G = 1$ and express masses in units of solar masses $M_\odot$.

\section{Numerical setup}
\label{sec:num_setup}

As demonstrated in previous studies \cite{Doulis:2022vkx,Doulis:2024aew,Doulis:2024ure}, the EFL scheme is applicable to any system of partial differential equations that can be expressed in conservative form, 
\begin{equation}
 \label{eq:cons_PDE}
 \partial_t \textbf{Q} + \partial_i \textbf{F}^i(\textbf{Q}) = \textbf{S},
\end{equation}
where the index $i$ runs over the spatial dimensions, $\textbf{Q}$ denotes the vector of conserved variables, $\textbf{F}^i$ are the physical fluxes along the spatial directions and $\textbf{S}$ is the vector of the sources. Within the EFL framework, the numerical fluxes obtained from the spatial discretization of \eqref{eq:cons_PDE} are formulated as a superposition of a high-order (HO) flux $f^{\,\mathrm{HO}}$, which may become unstable in non-smooth regions, and a low-order (LO) flux $f^{\,\mathrm{LO}}$ that provides the required robustness. The entropy production of the system is used to regulate the transition between the two fluxes through a continuous function $\theta \in [0,1]$ entering the flux-limiter:
\begin{equation}
 \label{eq:flx_lim}
  f_{i\pm1/2} = \theta_{i\pm1/2} \, f^{\,\mathrm{HO}}_{i\pm1/2} + 
  (1-\theta_{i\pm1/2}) \, f^{\,\mathrm{LO}}_{i\pm1/2},
\end{equation}
where all quantities are evaluated at the cell interfaces. 

The flux $f^{\,\mathrm{HO}}$ is constructed using the Rusanov (local Lax–Friedrichs) flux-splitting approach with reconstruction performed on the characteristic fields \cite{Mignone:2010br,Bernuzzi:2016pie}. A fifth-order unfiltered central stencil (CS5) is always used for reconstruction. The flux $f^{\,\mathrm{LO}}$ is computed using the local Lax-Friedrichs (LLF) central scheme with reconstruction performed directly on the primitive variables \cite{Thierfelder:2011yi}. Primitive variable reconstruction is performed with the fifth-order weighted-essentially-non-oscillatory finite difference scheme WENOZ \cite{Borges:2008a}. 

The weighting function $\theta$ is evaluated dynamically from the local entropy production \cite{Doulis:2022vkx,Doulis:2024aew,Doulis:2024ure}. For the piecewise-polytropic equations of state employed in this work, the specific entropy is given by
\begin{equation}
 \label{eq:entr}
  s = \ln \left( \frac{p}{\rho^\Gamma} \right),
\end{equation}
where $\rho$ denotes the rest-mass density, $p$ the pressure and $\Gamma$ the adiabatic index.

In the present work, we consider a number of hybrid EoSs, namely APR \cite{Akmal:1998cf}, ALF2 \cite{Alford:2004pf}, SLy \cite{Douchin:2001sv}, H4 \cite{Glendenning:1991es} and MPA1 \cite{Muther:1987xaa}. These models provide a broad sampling of NS matter properties commonly used in numerical relativity simulations. In the hybrid formulation \cite{Shibata:2005ss}, the pressure is decomposed into a cold component, obtained from a piecewise-polytropic approximation to one of the zero-temperature EoSs listed above, and an additional thermal contribution that accounts for shock heating and other non-isentropic processes. The thermal pressure is modelled using a $\Gamma$-law EoS with $\Gamma = 1.75$. 

In simulations employing hybrid EoSs, artificial heating manifests itself as an unphysical growth of the thermal component of the internal energy. Since the cold contribution depends only on the local density, such effects are also reflected in the evolution of the total specific internal energy $\epsilon$, which we therefore adopt as the primary diagnostic of numerical heating.

All simulations presented here are evolved with the general relativistic hydrodynamics code BAM \cite{Bruegmann:2006ulg,Thierfelder:2011yi,Dietrich:2015iva,Bernuzzi:2016pie}. The EFL method has been fully integrated into BAM and implemented as part of its core numerical infrastructure \cite{Doulis:2022vkx}.

BAM does not evolve the temperature or composition explicitly, but instead advances the specific internal energy as part of the primitive variable set. As a consequence, fully finite-temperature EoSs, which require the evolution or inversion of the temperature and composition, are not currently supported. Instead, thermal effects are modelled using hybrid EoSs, which provide a numerically robust approximation suitable for this framework. 

Recent developments have extended BAM to allow the use of finite-temperature, microphysical EoSs \cite{Gieg:2022mut,Schianchi:2023uky}. However, even in this extended framework, the temperature is not evolved as an independent dynamical variable. Instead, thermal effects are incorporated by reconstructing the temperature from the evolved rest-mass density, internal energy and electron fraction through tabulated EoS inversions. While this approach enables the inclusion of temperature-dependent microphysics, it differs from a fully self-consistent temperature evolution. Moreover, the numerical fluxes employed in these simulations do not make use of the EFL technique. As a result, the finite-temperature extension of BAM cannot presently be used to assess the impact of the EFL method on artificial surface heating. Consequently, the use of hybrid EoSs currently provides the only viable framework to include thermal effects in the investigations carried out in this work.  

Vacuum regions are treated by introducing a static, low-density, cold atmosphere surrounding the NS \cite{Thierfelder:2011yi}. The atmosphere density is prescribed as
\begin{equation} 
 \label{eq:atmos} 
  \rho_{\rm atm} = f_{\rm atm}\, \mathrm{max}\, \rho(t = 0), 
\end{equation} 
where $\mathrm{max}\, \rho(t = 0)$ denotes the initial maximum rest-mass density.
During the evolution, grid cells with rest-mass density falling below the threshold $\rho_{\rm thr} = f_{\rm thr}\rho_{\rm atm}$ are reset automatically to $\rho_{\rm atm}$.

In the following, the results obtained with the EFL scheme are compared with the ``hybrid'' HO-LLF algorithm \cite{Bernuzzi:2016pie}. The latter employs the high-order HO-WENOZ scheme \cite{Bernuzzi:2016pie} above a density threshold $\rho_\mathrm{hyb}$, this is the high-order scheme that we use to approximate the HO flux $f^{\,\mathrm{HO}}$ in \eqref{eq:flx_lim}, but with WENOZ instead of CS5. Below $\rho_\mathrm{hyb}$, the method switches to the standard second-order LLF-WENOZ scheme, which uses LLF for the numerical fluxes and WENOZ for the reconstruction of the primitive variables \cite{Thierfelder:2011yi}.

\section{Isolated neutron stars}
\label{sec:sns}

Tolman-Oppenheimer-Volkoff (TOV) initial data are constructed for a set of cold, spherically symmetric, non-rotating NSs in hydrostatic equilibrium described by the hybrid EoSs of \autoref{tab:tov_sim}. For each EoS, the central rest-mass density is specified and the TOV equations are solved to obtain equilibrium configurations with gravitational masses in the range $M \simeq 1.1-1.5 M_\odot$, as summarized in \autoref{tab:tov_sim}. The corresponding baryonic masses and stellar radii are determined self-consistently from the TOV solutions. These models span a representative range of NS masses, radii and compactness.  

\begin{table}[h]
 \centering    
 \caption{Initial single neutron star configurations. Columns: EoS, gravitational mass $M$, baryonic mass $M_b$, radius $R$, central rest-mass density $\rho_c$ and compactness $C$.}
 \begin{tabular}{cccccc}        
  \hline
  \hline
  EoS  & $M\,[M_\odot]$ & $M_b\,[M_\odot]$ & $R\,[\mathrm{km}]$ & $\rho_c\,[\times10^{-3}]$ & $C$ \\
  \hline
  H4   & 1.288 & 1.396 & 13.548 & 0.850 & 0.140 \\    
  APR4 & 1.074 & 1.166 & 10.983 & 1.250 & 0.144 \\  
  SLy  & 1.123 & 1.220 & 11.456 & 1.200 & 0.145 \\
  ALF2 & 1.409 & 1.560 & 12.389 & 1.080 & 0.168 \\  
  MPA1 & 1.459 & 1.625 & 12.166 & 1.120 & 0.177 \\  
  \hline
  \hline
 \end{tabular}
 \label{tab:tov_sim}
\end{table}

The computational domain consists of three fixed refinement levels, $l=(0,1,2)$, arranged in a nested-box configuration. Simulations are performed with 160 grid points per direction leading to grid spacings $h=(0.45,0.225,0.1125)$ for successive refinement levels. It is ensured that the NS is entirely covered by the finest box throughout the evolution. Each TOV configuration is evolved for approximately 25ms. This duration is sufficient to capture secular numerical effects such as artificial heating. Radiative (absorbing) boundary conditions are used in all single star simulations. The spacetime is dynamically evolved using the BSSNOK formulation. Octant symmetry is employed to reduce computational costs.

\begin{figure}[h]
 \includegraphics[width=0.49\textwidth]{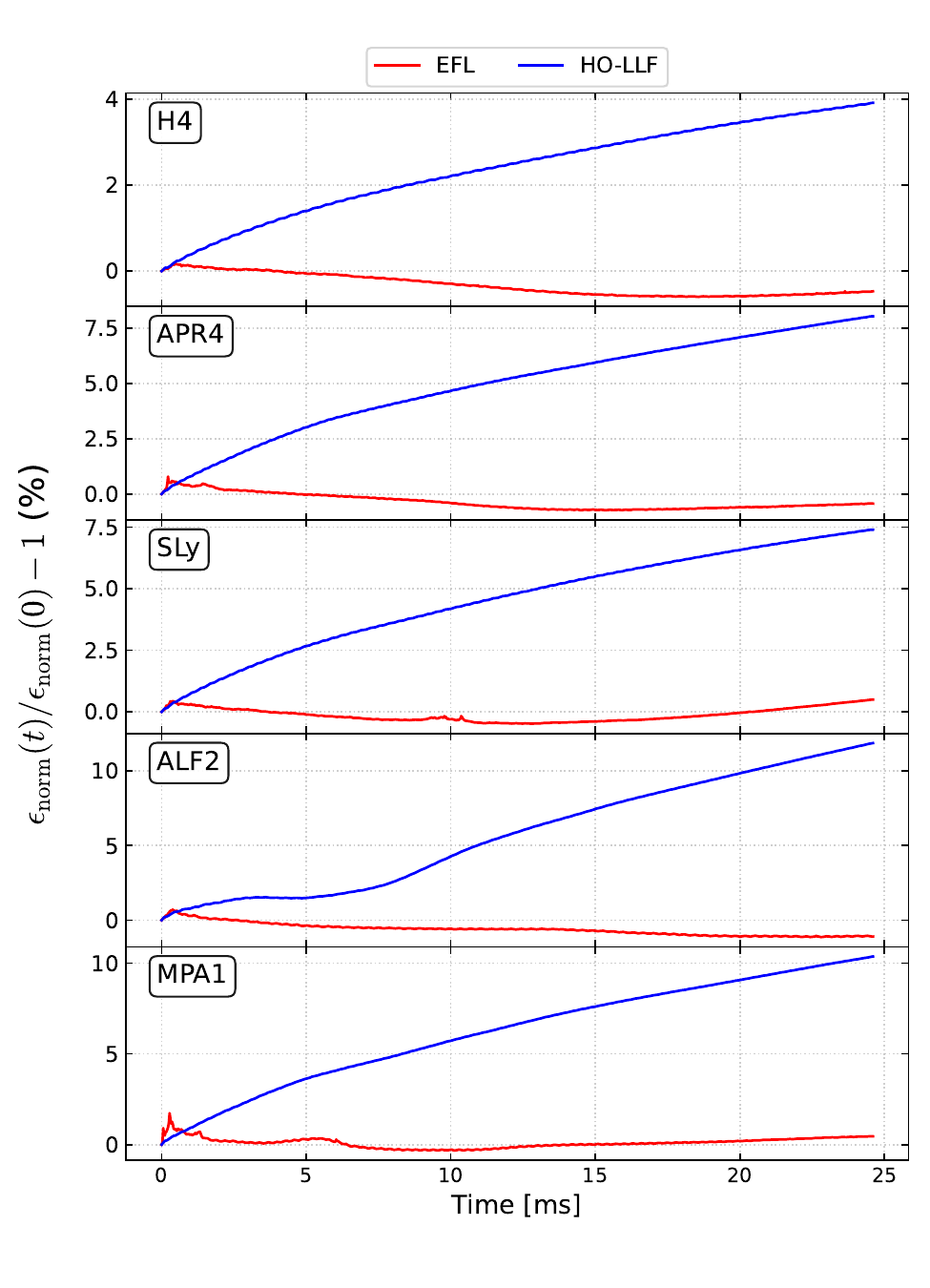}
 \caption{Time evolution of the $L_2$-norm of the specific internal energy for isolated NS simulations employing different hybrid EoSs. Results obtained with the HO-LLF scheme are compared to those obtained using the EFL method.}
 \label{fig:tov_eps_norm}
\end{figure}

\begin{figure}[h]
 \includegraphics[width=0.49\textwidth]{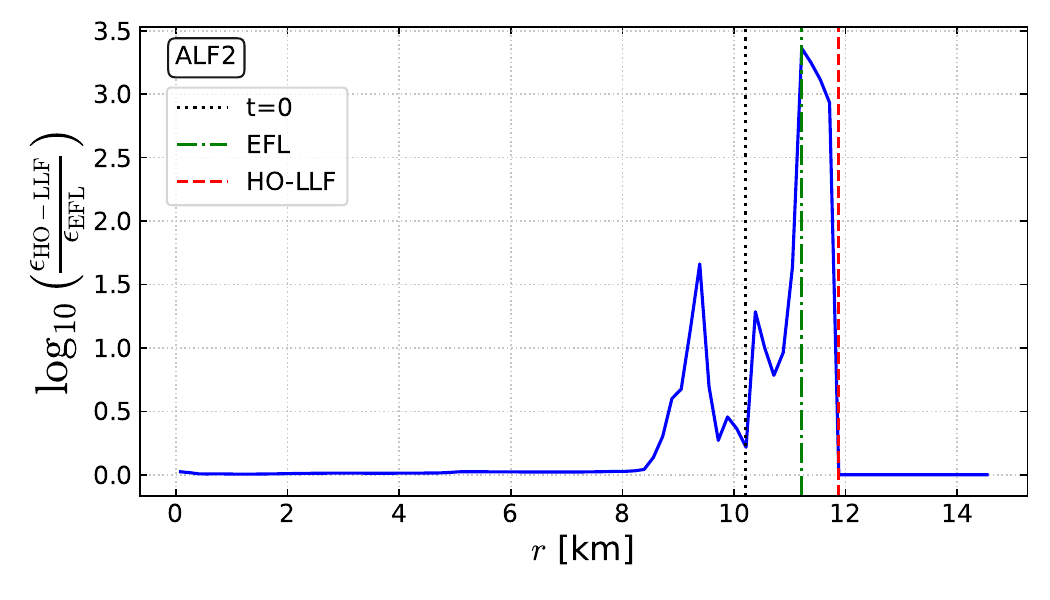}
 \caption{Radial profile of the ratio $\epsilon_\mathrm{HO-LLF} / \epsilon_\mathrm{EFL}$ for the ALF2 TOV model at time 25 ms. The horizontal axis shows the isotropic coordinate radius used by BAM. The black dotted line marks the initial stellar surface, while the red dashed and green dash-dotted lines indicate the stellar surface at 25ms for the HO-LLF and EFL schemes, respectively. The stellar surface is defined as the location where the rest-mass density falls to the atmosphere floor.}
 \label{fig:tov_eps_ratio}
\end{figure}

\autoref{fig:tov_eps_norm} shows the temporal evolution of the relative variation of the $L_2$-norm of the specific internal energy for simulations employing the hybrid EoSs of \autoref{tab:tov_sim}. Results obtained using the HO-LLF method are compared against those obtained with the EFL scheme. Across all EoSs, the HO-LLF scheme exhibits a systematic and monotonic increase of the internal energy norm, reaching deviations of up to several percent over a 25 ms evolution, indicating a persistent artificial surface heating pattern. In contrast, simulations performed with the EFL method show a markedly improved behaviour, with the relative variation of the internal energy remaining close to zero and typically bounded within sub-percent levels throughout the evolution. This trend is observed consistently across all stellar models, despite differences in mass, radius and compactness, demonstrating that the mitigation of spurious heating achieved by the EFL scheme is robust with respect to the underlying EoS.

The small negative relative variations observed in several EFL evolutions should not be interpreted as physical cooling, but rather as a consequence of the numerical relaxation of the discretized initial data towards the equilibrium supported by the numerical scheme. Discretization and interpolation errors can cause the initial configuration to deviate slightly from the corresponding discrete equilibrium, resulting in a small reduction of the $L_2$-norm during the evolution. These variations remain small and bounded, in contrast to the systematic secular increase observed with the HO-LLF scheme.

\begin{figure}[h]
 \includegraphics[width=0.49\textwidth]{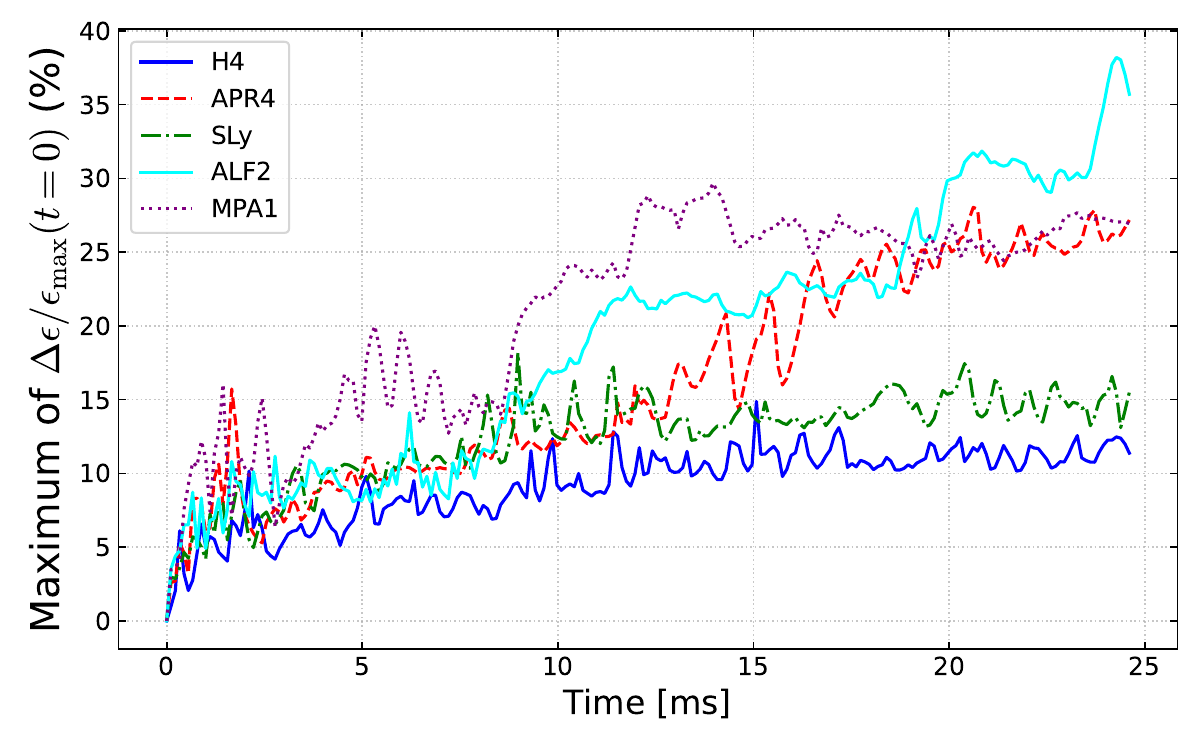}
 \caption{Time evolution of the maximum value of the difference $\Delta \epsilon = \epsilon_\mathrm{HO-LLF} - \epsilon_\mathrm{EFL}$ between the HO-LLF and EFL schemes for all considered EoSs.}
 \label{fig:tov_eps_diff_max}
\end{figure}

\begin{figure}[h]
 \includegraphics[width=0.49\textwidth]{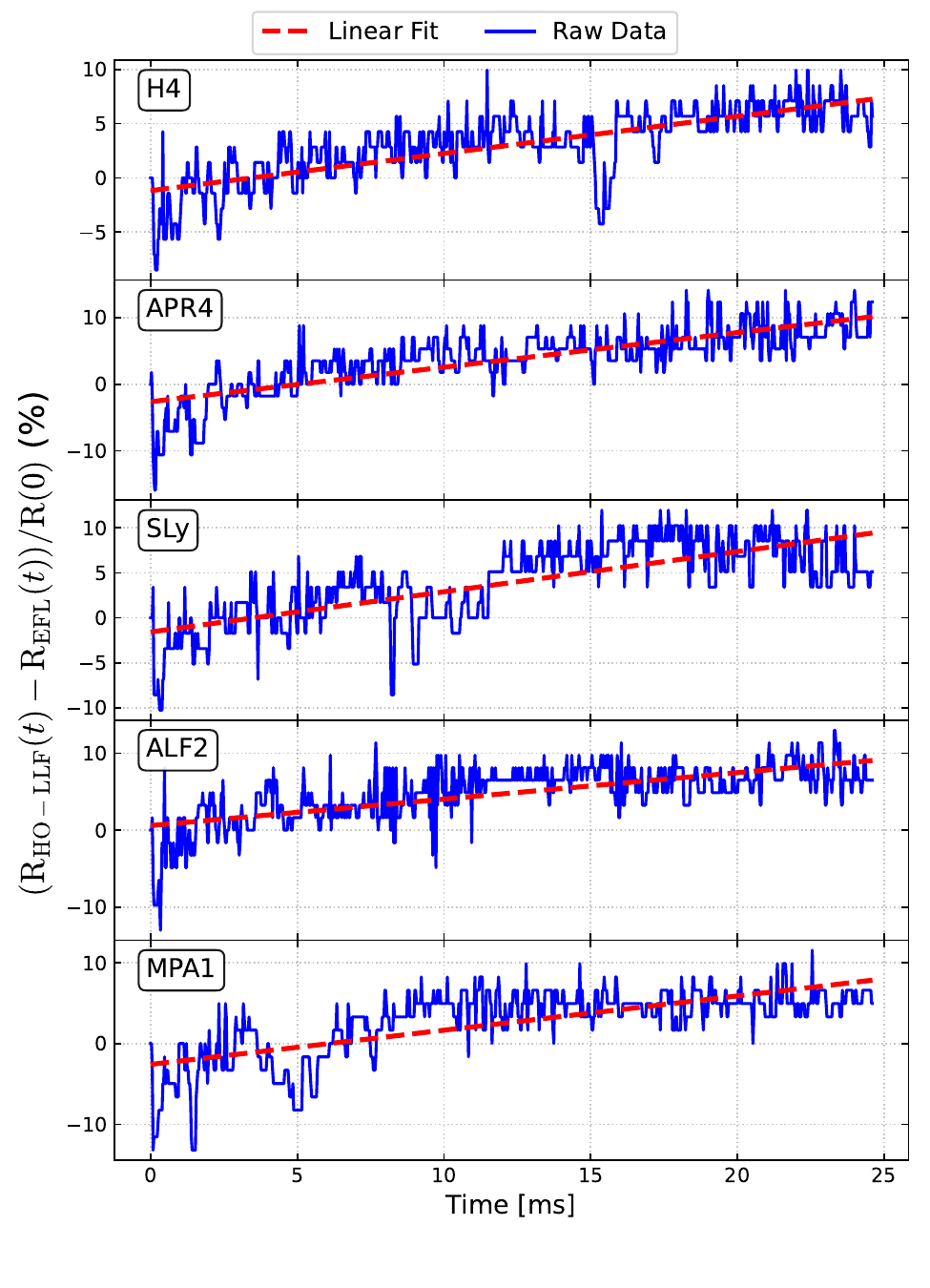}
 \caption{Time evolution of the difference in the stellar radius between HO-LLF and EFL simulations for the TOV models considered in this work. Solid blue lines show the raw data, while dashed red lines denote linear least-squares fits highlighting the secular trend.}
 \label{fig:tov_radius_diff}
\end{figure}

\begin{figure}[h]
 \includegraphics[width=0.49\textwidth]{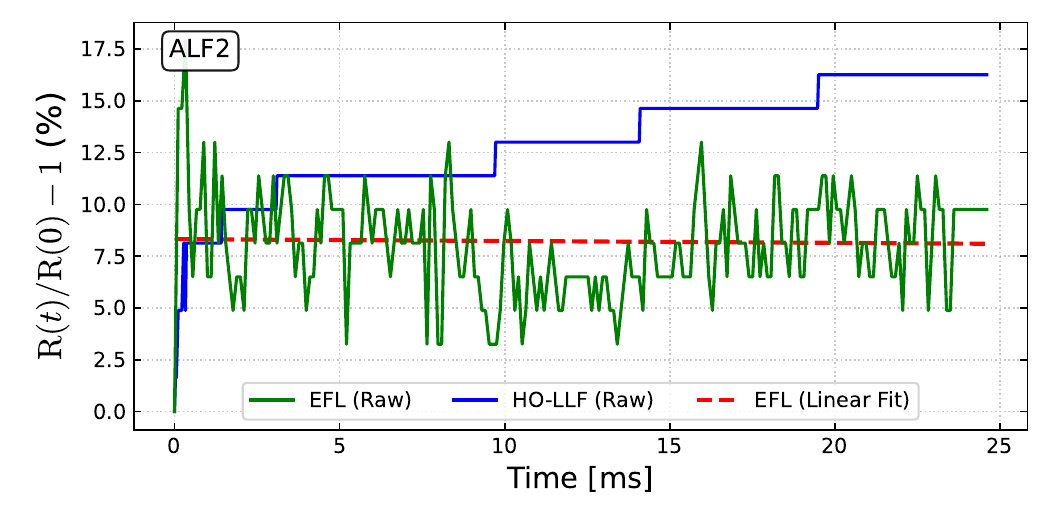}
 \caption{Time evolution of the stellar radius for the ALF2 TOV model. The radius is shown for simulations performed with the HO-LLF (solid blue line) and EFL (solid green line) numerical flux schemes. Raw data are plotted with solid lines, while a linear least-squares fit to the EFL data is shown with a dashed red line.}
 \label{fig:tov_radius}
\end{figure}

\autoref{fig:tov_eps_ratio} illustrates the radial profile of the logarithmic ratio of the specific internal energy for the HO–LLF and EFL schemes, i.e. $\log_{10}(\epsilon_\mathrm{HO-LLF} / \epsilon_\mathrm{EFL})$, for the ALF2 TOV model\footnote{A qualitatively similar behaviour is observed for the other TOV models of \autoref{tab:tov_sim}.} at $t=25$ ms. Throughout most of the stellar interior, the difference between the two schemes remains small, indicating good agreement in regions where the flow is smooth. In contrast, a pronounced and spatially localized peak develops near the stellar surface. This behaviour indicates that the excess internal energy produced by the HO-LLF scheme is generated primarily at the stellar boundary. In particular, this strong surface-localized enhancement provides a clear explanation for the secular growth of the $L_2$-norm of the specific internal energy observed in \autoref{fig:tov_eps_norm} when the numerical fluxes are computed with the HO-LLF scheme. As the simulation proceeds, this surface-generated excess internal energy is advected away from the surface into the stellar interior and redistributed throughout the star, leading to the monotonic increase of the global energy norm. The enhanced heating at the stellar surface leads to an outward displacement of the stellar surface during the evolution. As indicated by the positions of the vertical lines in \autoref{fig:tov_eps_ratio}, this displacement is significantly smaller for the EFL than for the HO-LLF scheme. This observation is consistent with the results presented in \autoref{fig:tov_radius}.

Following the spatial analysis of the ratio $\epsilon_\mathrm{HO-LLF} / \epsilon_\mathrm{EFL}$ in \autoref{fig:tov_eps_ratio}, we now study the temporal evolution of the global maximum of $\Delta \epsilon = \epsilon_\mathrm{HO-LLF} - \epsilon_\mathrm{EFL}$ for all considered EoSs in \autoref{fig:tov_eps_diff_max}. For each model, we track the most extreme local deviation between the HO-LLF and EFL schemes by plotting the maximum value of $\Delta \epsilon$ over the spatial domain at each time step. Across all EoSs, the maximum  difference exhibits a rapid initial growth during the early transient phase, followed by a sustained increase over the duration of the simulation. When interpreted together with the radial profiles shown in \autoref{fig:tov_eps_ratio}, this behaviour indicates that the growing extrema are associated with the persistent surface-localized excess internal energy generated by the HO-LLF scheme. The continued growth of the maximum of $\Delta \epsilon$ demonstrates that the surface-generated numerical heating persists and intensifies over time.

\begin{figure*}[th]
    \centering
    \includegraphics[width=0.99\textwidth]{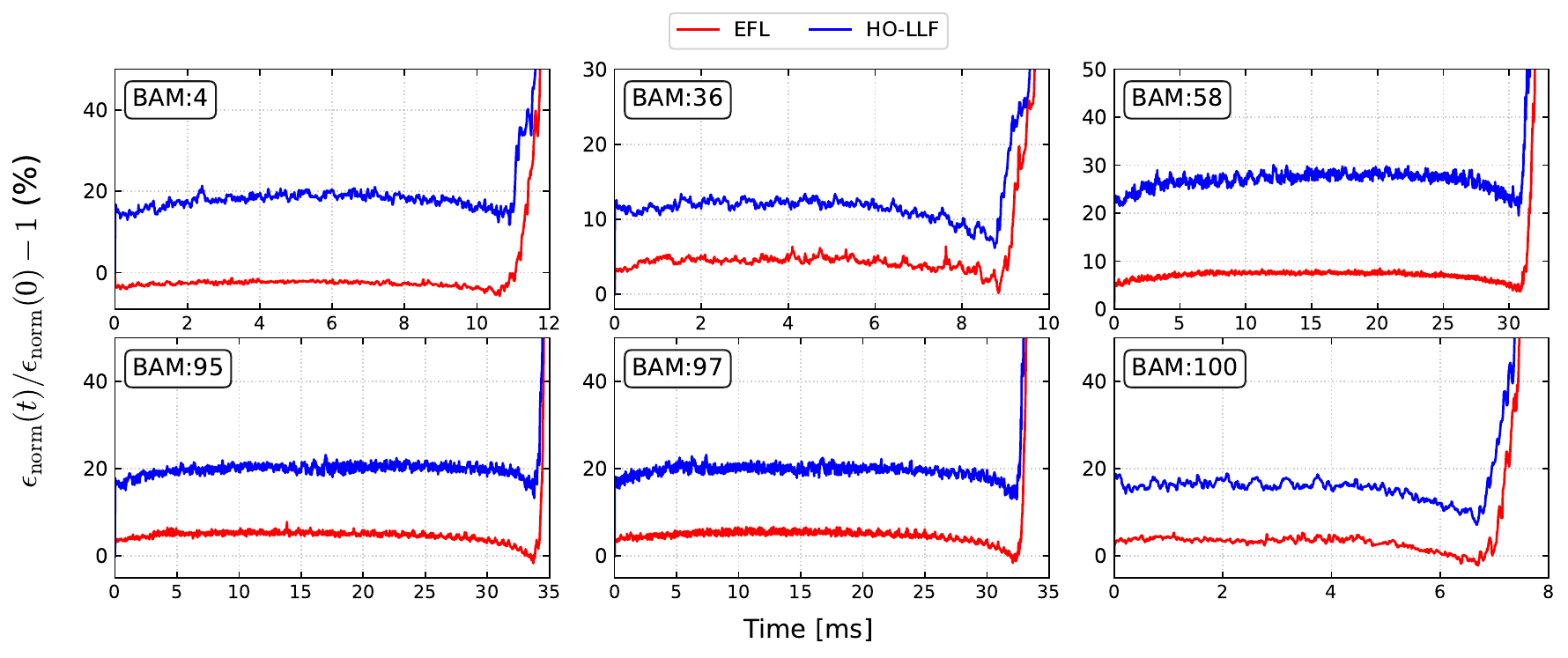}
    \caption{Time evolution of the $L_2$-norm of the specific internal energy for the BNS simulations of \autoref{tab:bns_sim}. Results obtained with the HO-LLF (blue solid line) scheme are compared to those obtained using the EFL (red solid line) method. Each panel corresponds to a different binary configuration. The vertical rise at the end of each evolution marks the onset of merger.}
    \label{fig:bns_eps_norm}
\end{figure*}

In \autoref{fig:tov_radius_diff}, we show the temporal evolution of the difference in the stellar radius between simulations performed with the HO-LLF and EFL schemes for each considered EoS. A linear least-squares fit (dashed red line) is applied to the raw data (solid blue line) to highlight the long-term trend. For all EoSs considered, the radius difference displays a systematic positive drift over time, indicating a progressively larger stellar radius in the HO-LLF evolutions. Despite short-time fluctuations, the late-time discrepancy reaches values of order $\sim$10\%. The origin of this trend can be traced back to \autoref{fig:tov_radius}, where the temporal evolution of the stellar radius obtained with both HO-LLF and EFL schemes is presented for the ALF2 EoS.\footnote{A qualitatively similar behaviour is observed for the other EoSs.} It is apparent that the HO-LLF scheme leads to a continuous, step-like growth of the NS radius, whereas the EFL evolution, after an initial rapid adjustment, gradually stabilizes and maintains the stellar radius at approximately 7.5\% above its initial value. This behaviour is consistent with the excess internal energy generated by the HO-LLF scheme, as shown in \autoref{fig:tov_eps_norm}--\autoref{fig:tov_eps_diff_max}. The spurious increase in internal energy provides additional pressure support in the outer layers of the star, leading to the observed artificial expansion of the stellar radius. In all cases the EFL method exhibits a systematically smaller increase of the stellar radius throughout the evolution, indicating a significant reduction of artificial surface heating compared to the standard HO-LLF scheme.

\section{Binary neutron stars}
\label{sec:bns}

Building on the results obtained in the single star case, we now extend the analysis to BNS simulations. While the single star setup provides a controlled environment to isolate numerical heating effects, the binary case allows us to assess their impact in a fully dynamical and strongly interacting spacetime. This enables us to investigate whether the  behaviour observed in isolated stars persists in more realistic astrophysical scenarios. In the following, we analyse a representative set of binaries to inspect whether the use of the EFL scheme mitigates artificial surface heating during dynamical inspirals.

The BNS configurations \cite{Dietrich:2018phi,Gonzalez:2022mgo} considered in this work are summarized in \autoref{tab:bns_sim}. The selected dataset spans a representative range of binary configurations employing the ALF2, H4, MPA1 and SLy EoSs, thereby covering NS matter models with different stiffness, compactness and tidal deformability properties. All binaries are equal mass, irrotational systems with total gravitational masses close to $M \simeq 2.7,M_\odot$. The sample includes configurations with different initial orbital separations and gravitational wave frequencies, corresponding to binaries undergoing approximately $3$--$11$ orbits prior to merger. In particular, \texttt{BAM:100} starts from a significantly higher initial frequency and therefore undergoes fewer inspiral cycles than the other models. All configurations are constructed using LORENE initial data \cite{Gourgoulhon:2000nn}, except for \texttt{BAM:95}, which employs the SGRID solver \cite{Tichy:2012rp}, allowing us to assess the robustness of the results with respect to differences in the initial data construction. Overall, this set of simulations provides a diverse set of inspiral conditions for investigating the impact of the EFL scheme on artificial heating effects in BNS simulations.

\begin{table}[h]
 \centering    
 \caption{Initial BNS configurations. Columns: simulation name, EoS, number of orbits $N_{\rm orb}$, total gravitational mass $M\,[M_\odot]$, total baryonic mass $M_b\,[M_\odot]$, ADM mass $M_{\rm ADM}\,[M_\odot]$, ADM angular momentum $J_{\rm ADM}\,[M_\odot^2]$ and gravitational wave frequency $M\omega_{22}$. All
configurations are equal mass and irrotational.}
 \begin{tabular}{cccccccc}
  \hline
  \hline
  Name & EoS & $N_{\rm orb}$ & $M$ & $M_b$ & $M_{\rm ADM}$ & $J_{\rm ADM}$ & $M\omega_{22}$ \\
  \hline
  \texttt{BAM:4}   \cite{Dietrich:2015iva} & ALF2 & 6  &2.702  & 2.978 & 2.675 & 7.148 & 0.052 \\
  \texttt{BAM:36}  \cite{Dietrich:2015iva} & H4   & 5  & 2.701 & 2.939 & 2.674 & 7.133 & 0.052 \\
  \texttt{BAM:58}  \cite{Bernuzzi:2014kca} & MPA1 & 10 & 2.700 & 2.979 & 2.678 & 7.659 & 0.038 \\
  \texttt{BAM:95}  \cite{Dietrich:2017aum} & SLy  & 11 & 2.700 & 2.989 & 2.678 & 7.686 & 0.038 \\
  \texttt{BAM:97}  \cite{Bernuzzi:2014kca} & SLy  & 11 & 2.700 & 2.989 & 2.678 & 7.658 & 0.038 \\
  \texttt{BAM:100} \cite{Bernuzzi:2016pie} & SLy  & 3  & 2.700 & 2.989 & 2.671 & 6.872 & 0.060 \\
  \hline
  \hline
 \end{tabular}
 \vspace{0.5em}
 
 {\footnotesize
 \textit{Note:}
 \texttt{BAM:95} and \texttt{BAM:97} employ different initial data solvers, namely SGRID and LORENE, respectively.
 }
 \label{tab:bns_sim}
\end{table}

The computational domain is the same for all BNS simulations and consists of seven fixed refinement levels, $l=(0,\dots,6)$, arranged in a nested-box configuration. Simulations are performed with 128 grid points per direction, resulting in a coarsest grid spacing of $h_0=7.296$, with the resolution increasing by a factor of two between successive refinement levels, i.e. $h=(7.296,\dots,0.114)$. The metric is evolved with the Z4c scheme. Bitant symmetry and standard radiative boundary conditions are employed for all BNS simulations.

\begin{figure*}[th]
 \includegraphics[width=0.32\textwidth]{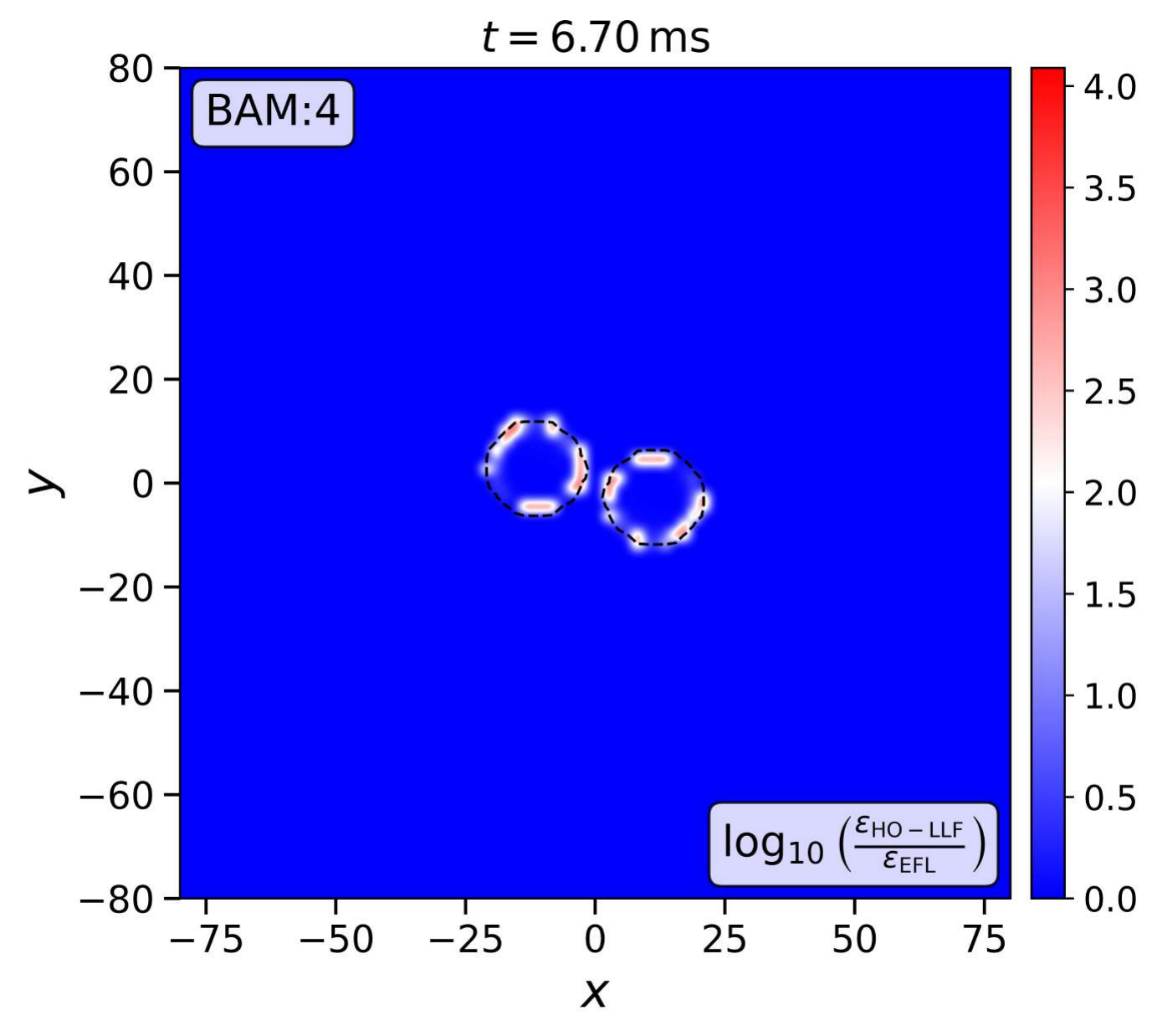}
 \includegraphics[width=0.32\textwidth]{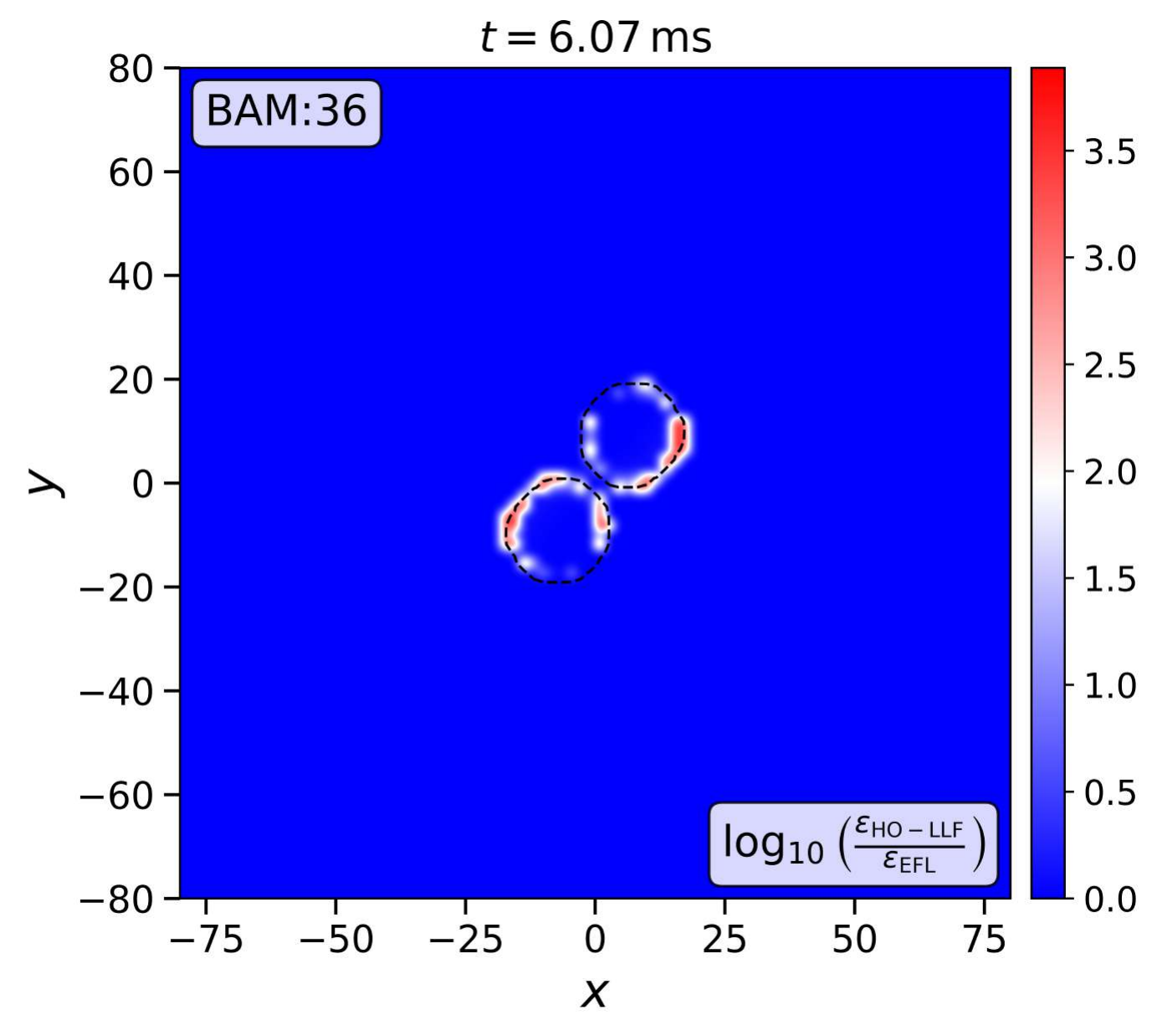}
 \includegraphics[width=0.32\textwidth]{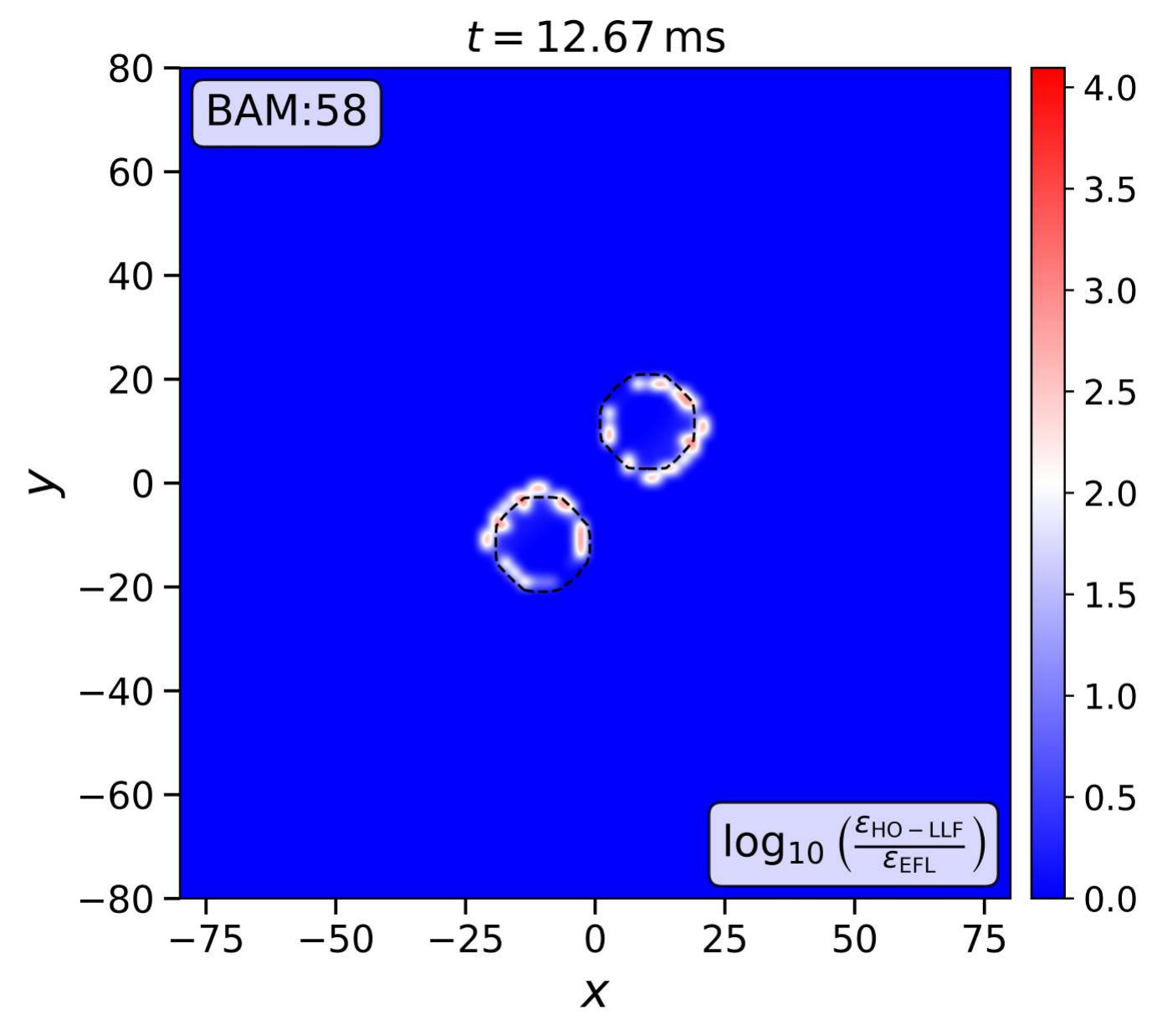}\\
 \includegraphics[width=0.32\textwidth]{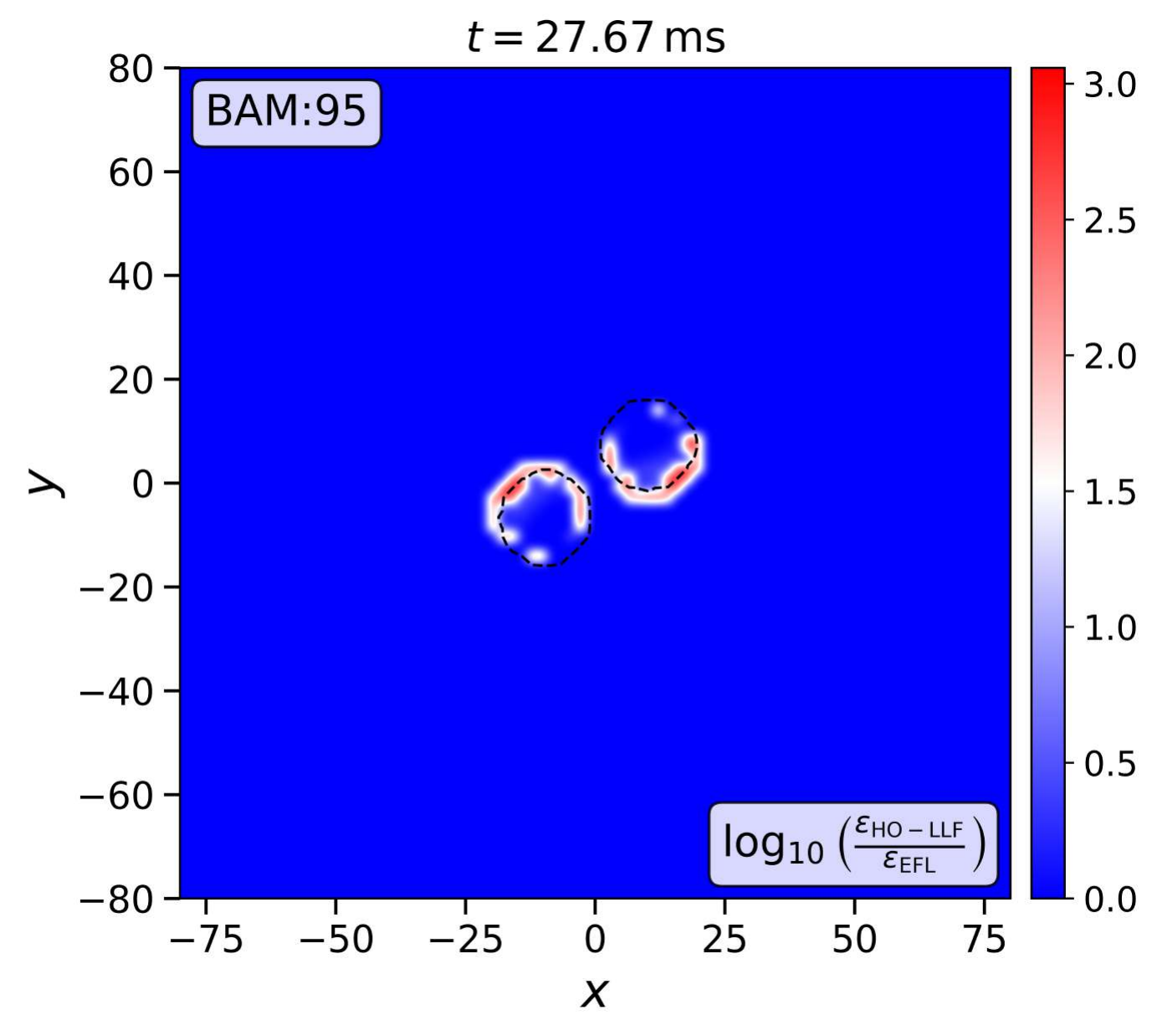}
 \includegraphics[width=0.32\textwidth]{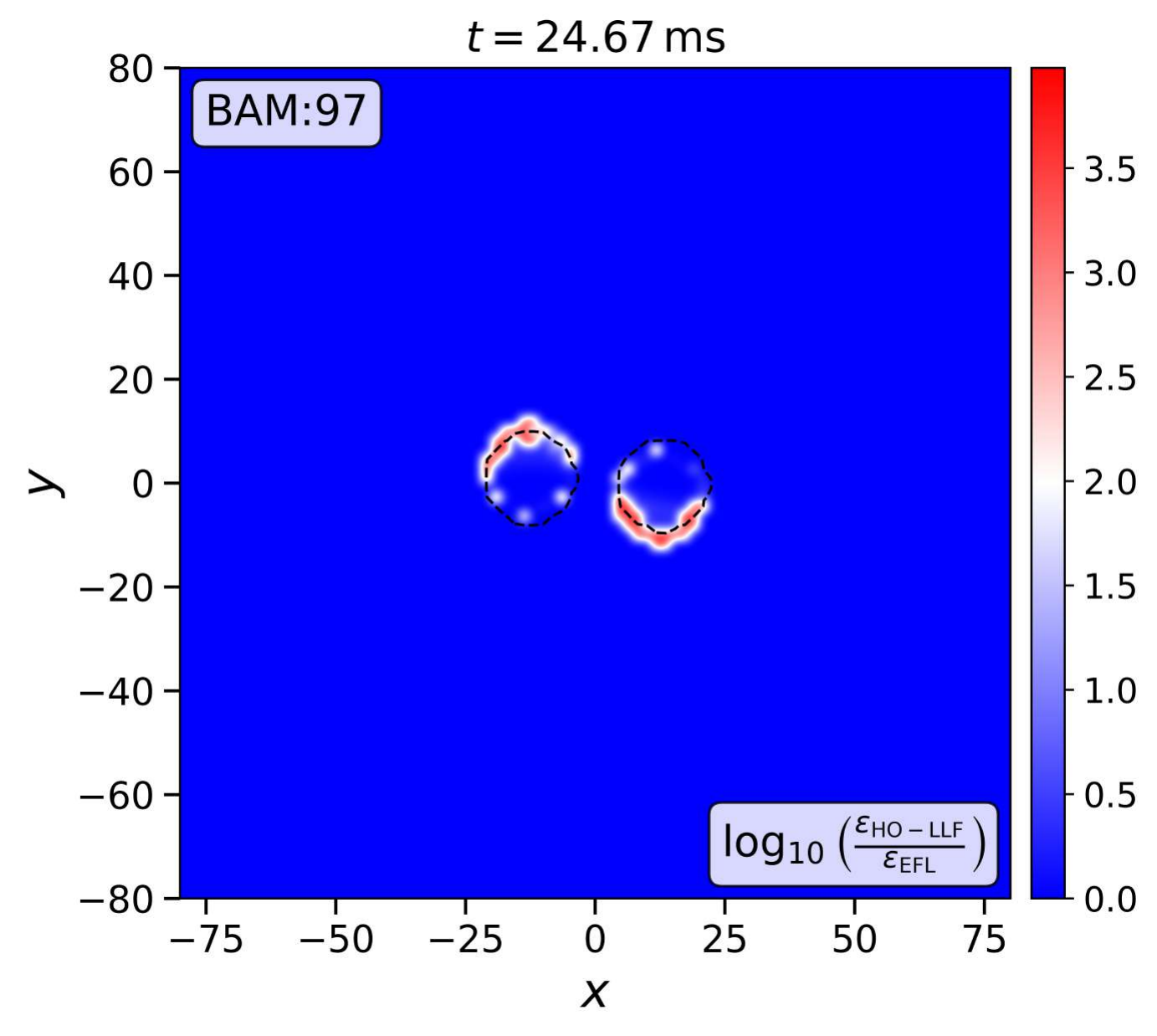}
 \includegraphics[width=0.32\textwidth]{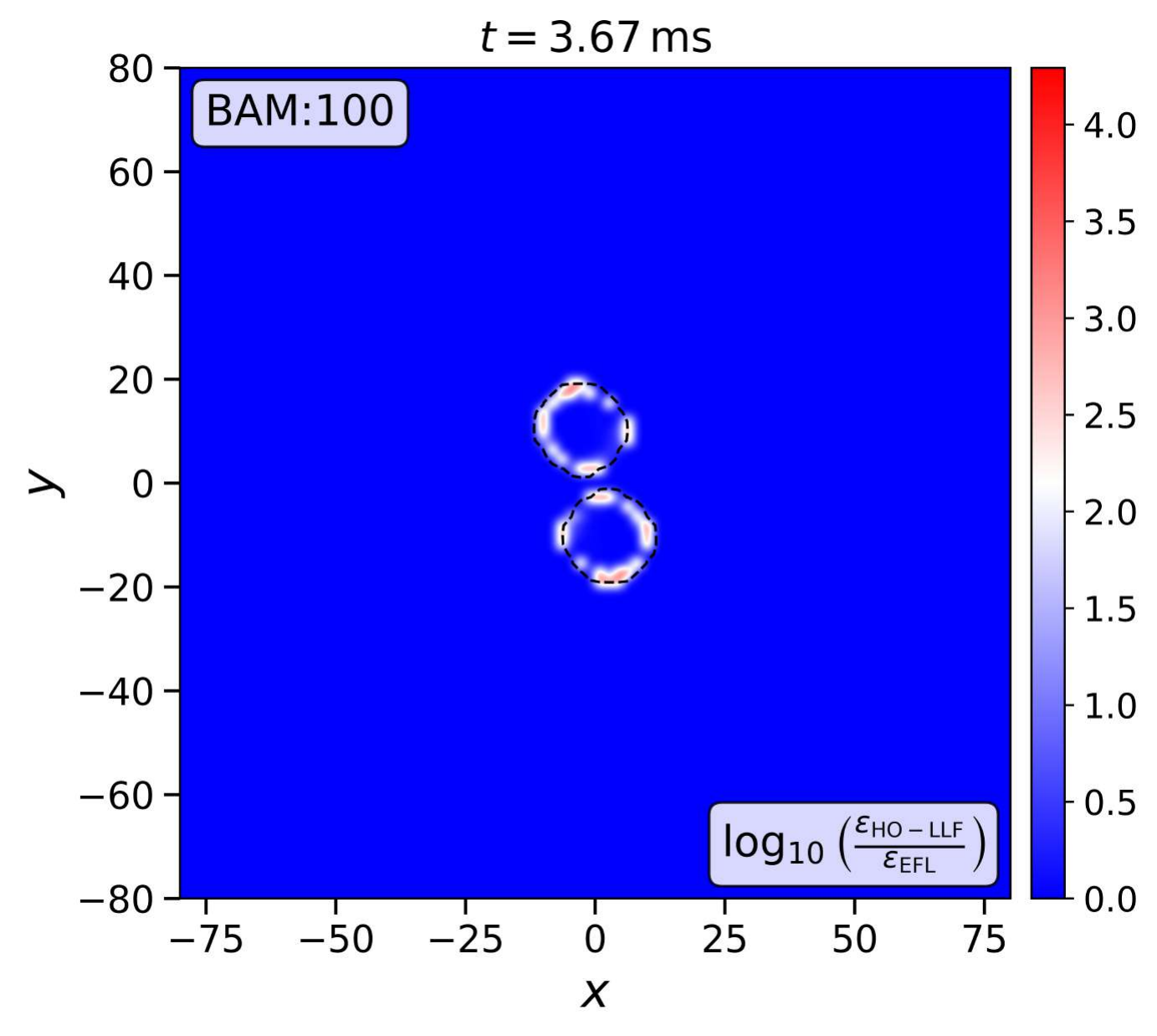}
 \caption{2-dimensional spatial distribution of the ratio $\epsilon_{\rm HO-LLF}/\epsilon_{\rm EFL}$ for the BNS configurations listed in \autoref{tab:bns_sim} at representative times during the late inspiral. The ratio highlights the regions where the HO-LLF scheme produces larger specific internal energy than the EFL scheme. Each panel corresponds to a different binary configuration.}
 \label{fig:bns_eps_ratio}
\end{figure*}

The time evolution of the relative variation of the $L_2$-norm of the specific internal energy for the BNS configurations listed in \autoref{tab:bns_sim} is presented in \autoref{fig:bns_eps_norm}. As in the isolated NS evolutions, see \autoref{sec:sns}, the results obtained with the HO-LLF scheme are compared with those obtained using the EFL method. For all binary configurations considered, the HO-LLF simulations exhibit a substantially larger increase of the internal energy norm throughout the inspiral than the corresponding EFL evolutions. After an initial transient, the HO-LLF curves remain systematically above the EFL ones, indicating the continuous accumulation of excess internal energy during the binary evolution. Although the dynamical interaction between the stars introduces larger temporal fluctuations than in the single-star case, the qualitative behaviour remains unchanged: the EFL scheme consistently suppresses the secular growth of the internal energy associated with artificial surface heating. Shortly before merger, all models display a rapid increase in the internal energy norm as strong hydrodynamical interactions and tidal deformation become dominant. Nevertheless, the EFL simulations maintain significantly lower values than the corresponding HO-LLF runs, demonstrating that the reduction of spurious heating observed in the TOV tests carries over to fully dynamical BNS inspirals. 

Notice that the small initial decrease observed for the EFL evolution of BAM:4 is consistent with the relaxation of the discretized initial data towards the equilibrium of the numerical scheme and remains bounded throughout the inspiral. In contrast, the HO-LLF scheme rapidly develops a positive secular increase, indicating that the natural relaxation is dominated by artificial surface heating.

\begin{figure*}[th]
 \includegraphics[width=0.99\textwidth]{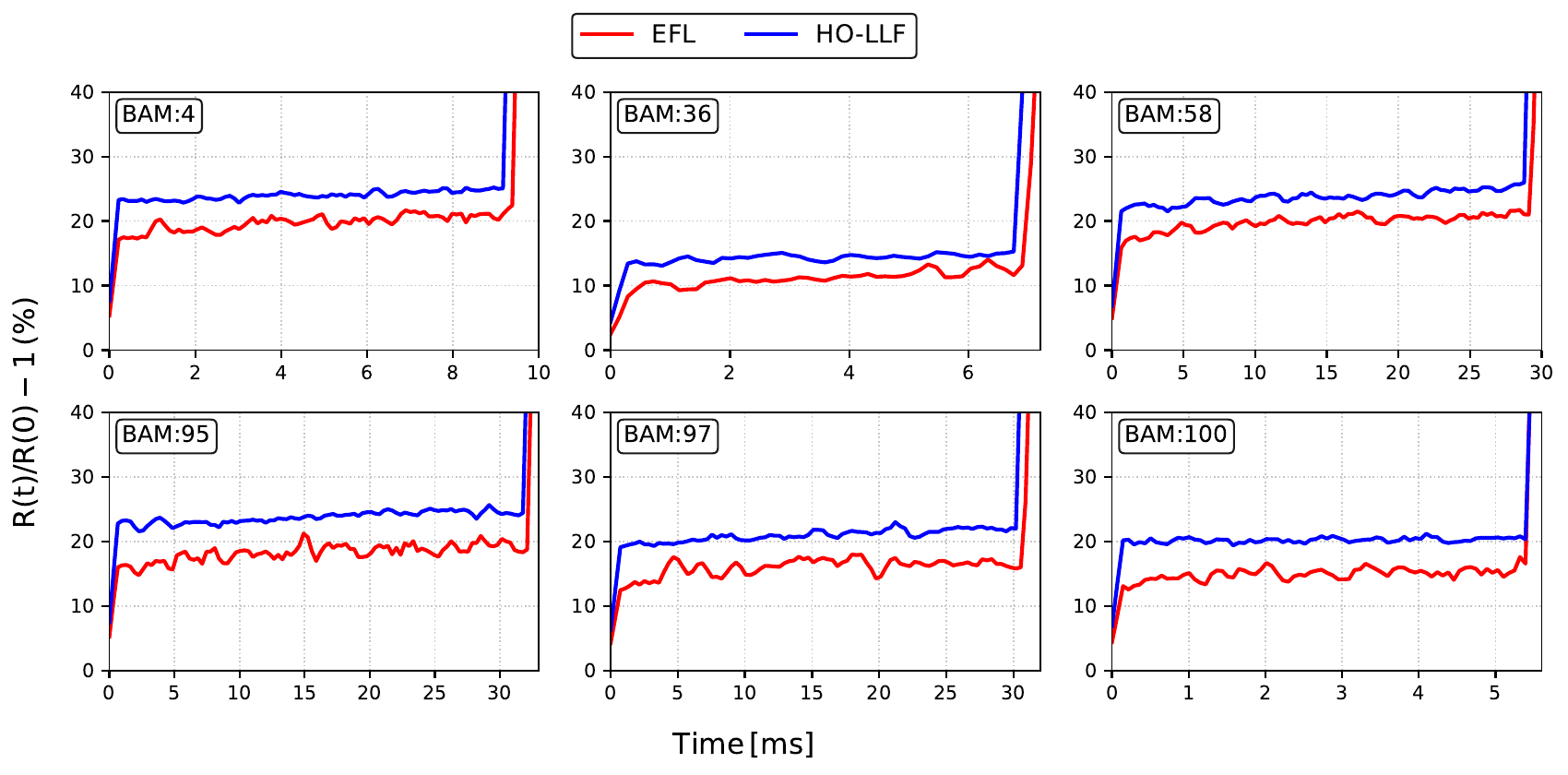}
 \caption{Time evolution of the stellar radius for the BNS simulations of \autoref{tab:bns_sim} evolved with the HO-LLF (blue solid line) and EFL (red solid line) schemes.}
 \label{fig:bns_radius}
\end{figure*}

Following the global diagnostics presented in \autoref{fig:bns_eps_norm}, \autoref{fig:bns_eps_ratio} illustrates the 2-dimensional spatial distribution of the logarithmic ratio $\log_{10}(\epsilon_\mathrm{HO-LLF}/\epsilon_\mathrm{EFL})$ for all BNS configurations of \autoref{tab:bns_sim} at representative times during the late inspiral. Consistent with the isolated NS results presented in \autoref{sec:sns}, the largest differences between the two schemes are localized at the NS surfaces, while the stellar interiors remain largely unaffected. The enhanced specific internal energy produced by the HO-LLF scheme therefore originates primarily in the low-density surface layers, where steep gradients and the artificial atmosphere make the numerical evolution particularly challenging. As the inspiral proceeds, this surface-generated excess energy is continuously replenished as the stars orbit and deform under tidal interactions, explaining the persistent secular increase of the global $L_2$-norm shown in \autoref{fig:bns_eps_norm}. The remarkably similar spatial pattern observed for all considered EoSs indicates that the suppression of the numerical heating at (and near) the stellar surface is a robust feature of the EFL scheme rather than a consequence of a particular stellar model or binary configuration.

The consequences of the excess surface heating on the stellar structure are illustrated in \autoref{fig:bns_radius}, which shows the temporal evolution of the relative stellar radius for all BNS configurations of \autoref{tab:bns_sim} evolved with the HO-LLF and EFL schemes. For each snapshot, the stellar surface is identified by the contour $\rho=\rho_{\rm surf}$, where $\rho_{\rm surf}=10^{-4}\rho_{\max}$ and $\rho_{\max}$ is the maximum density of the star at that time. The radius is then estimated as the mean distance from the stellar center, defined by the density maximum, to the corresponding isodensity contour. Consistent with the behaviour observed for isolated NSs (see \autoref{sec:sns}), the stellar radius remains systematically larger in the HO-LLF simulations throughout the inspiral. After a short initial relaxation phase, both schemes exhibit a gradual radius increase. However, the HO-LLF evolutions consistently show a greater expansion than the corresponding EFL runs, with differences of several percent persisting until shortly before merger. This behaviour is observed for all binary configurations considered, irrespective of the underlying EoS or initial orbital separation. The systematic radius increase is naturally explained by the artificial surface heating produced by the HO-LLF scheme. The excess specific internal energy generated near the stellar surface provides additional pressure support in the outer layers of the stars, leading to an artificial expansion that is largely suppressed when the EFL method is employed. The consistent reduction of the stellar radius obtained with EFL therefore provides further evidence that the mitigation of numerical heating translates into a measurable improvement in the global dynamics of BNS inspirals.

\section{Conclusions}
\label{sec:conclusions}

In this work, we have investigated the origin and evolution of artificial surface heating in NS simulations employing hybrid EoSs within the BAM code. Particular emphasis was placed on assessing the performance of the EFL scheme in reducing the spurious increase of the specific internal energy associated with the numerical treatment of the stellar surface. The analysis was carried out for both isolated NSs and BNS inspirals, allowing the robustness of the method to be evaluated in both controlled equilibrium configurations and fully dynamical systems.

For isolated NSs, see \autoref{sec:sns}, we demonstrated that the secular increase of the specific internal energy observed with the HO-LLF scheme originates primarily from the stellar surface. This artificial heating accumulates throughout the evolution, leading to a continuous increase of the global internal energy norm and producing an artificial expansion of the stellar radius. In contrast, the EFL scheme effectively suppresses this behaviour for all hybrid EoSs considered, maintaining the internal energy close to its initial value while substantially reducing the unphysical growth of the stellar radius. The EFL scheme reduces the accumulated increase in the internal energy by 4-12\%, depending on the EoS, and the associated radius expansion by approximately 10\%.

The same qualitative behaviour persists in BNS inspirals, see \autoref{sec:bns}. Despite the additional complexity introduced by orbital motion, tidal interactions and strong-field dynamics, the excess internal energy generated by the HO-LLF scheme remains localized near the NS surface and continuously accumulates during the inspiral. Simulations employing the EFL method consistently exhibit significantly lower internal energy, reduced surface heating and smaller stellar radii across all binary configurations considered, demonstrating that the improvement is robust with respect to the underlying EoS, initial orbital separation and initial data construction. The reduction in the internal energy growth and stellar expansion persists throughout the inspiral, with the EFL scheme reducing the internal energy growth by 8-20\%, depending on the EoS, and the stellar expansion by approximately 5\%. 

These results confirm and extend the findings of \cite{Gittins:2024jui}, showing that artificial surface heating is primarily a numerical effect associated with the treatment of steep gradients at the NS surface rather than an intrinsic feature of the physical evolution. Our results further demonstrate that this effect is not restricted to finite-temperature EoSs, but is also clearly present in simulations employing hybrid EoSs. The EFL method provides a simple and effective strategy for mitigating this numerical heating in NS simulations and represents a promising ingredient for future high-accuracy numerical relativity calculations.

The present work has focused on hybrid EoSs, for which the specific internal energy serves as an effective diagnostic of artificial heating. An important next step is to investigate the impact of the EFL method in simulations employing finite-temperature microphysical EoSs. To this end, the EFL method is currently being implemented in the GRMHD codes FIL (Frankfurt Illinois) and GRACE (General Relativistic Astrophysics Code for Exascale) \cite{Musolino:2026xms}, both of which evolve the temperature as a primitive variable. Extending the EFL method to finite-temperature evolutions within the finite-temperature extension of BAM \cite{Gieg:2022mut,Schianchi:2023uky} is left for future work. We expect that these developments will allow the impact of the EFL scheme to be assessed directly in simulations with realistic temperature-dependent microphysics and may provide a further step towards more accurate and physically reliable NS merger simulations.

\begin{acknowledgements}

We would like to thank Khalil Pierre for carefully proofreading the manuscript and for providing valuable comments and suggestions.
GD acknowledges funding from the European High Performance Computing Joint Undertaking (JU) and Belgium, Czech Republic, France, Germany, Greece, Italy, Norway, and Spain under grant agreement No 101093441 (SPACE).  
We acknowledge the EuroHPC JU for awarding us access to MareNostrum5 at BSC, Spain, through the Regular Access programme under project ID EHPC-REG-2025R02-175 and to Leonardo at CINECA, Italy, through the Development Access programme. The majority of the computations for this work were performed on these EuroHPC machines. Additional computations were performed on the CALEA and GOETHE clusters of Goethe University Frankfurt.

\end{acknowledgements}

%%______________________________________________________________


\end{document}